\documentclass[prl,twocolumn,aps,superscriptaddress,amsmath,amssymb,10pt]{revtex4-2}
\usepackage[utf8]{inputenc}
\usepackage[T1]{fontenc}

\usepackage{graphicx}
\usepackage{framed}
\usepackage{esvect}
\usepackage{amsmath,amssymb,color,bm}
\usepackage{mathrsfs}
\usepackage[dvipsnames]{xcolor}
\usepackage{siunitx}
\usepackage{hyperref}
\hypersetup{
	colorlinks=true,
	linkcolor=blue,
	citecolor=blue,
	filecolor=blue,
	urlcolor=blue,
	pdfstartview=FitH,
	pdfpagemode=UseNone
}

\newcommand\tb[1]{\textbf{#1}}

\newcommand\tn[1]{\textnormal{#1}}

\newcommand\mc[1]{\mathcal{#1}}

\newcommand\beq{\begin{equation}}
\newcommand\eeq{\end{equation}}
\newcommand\beqa{\begin{eqnarray}}
\newcommand\eeqa{\end{eqnarray}}

\def\w{\omega}

\def\H{\mc{H}}
\def\E{\mc{E}}

\begin{document}

\title{Crossing the Rotational Sound Barrier in a Quantum Solvent}

\author{Baptiste Coquinot}\email{Baptiste.Coquinot@ist.ac.at}
\affiliation{Institute of Science and Technology Austria (ISTA), Am Campus 1, 3400 Klosterneuburg, Austria}

\author{Giacomo Bighin}
\affiliation{University of Zagreb Faculty of Science, Department of Physics, Bijeni\v{c}ka 32, 10000 Zagreb, Croatia}


\author{Mikhail Lemeshko}
\affiliation{Institute of Science and Technology Austria (ISTA), Am Campus 1, 3400 Klosterneuburg, Austria}

\author{Ragheed Alhyder}
\affiliation{Institute of Science and Technology Austria (ISTA), Am Campus 1, 3400 Klosterneuburg, Austria}

\date{\today}

\begin{abstract}
Molecules embedded in superfluids provide an experimentally controllable platform for investigating impurity physics.
Here, we investigate a driven molecule rotating in a superfluid environment, including helium and Bose--Einstein condensates, at rotation frequencies similar to the bath dynamics. 
Within the experimentally relevant platform of an optical centrifuge, we show that the rotor remains localized up to a characteristic harmonic frequency that can realistically exceed the excitation energies of the bath, enabling access to ultrafast rotating impurities.
In the co-rotating frame, the bath excitations experience a rotational Doppler shift, generating angular-momentum-resolved resonances absent in equilibrium angulon theory. 
We identify a dissipative rotational sound barrier at which the molecule resonantly emits bath excitations and undergoes strong angular momentum exchange with the surrounding medium. 
Overall, we establish the dynamical phase diagram of the driven rotor in a quantum solvent and introduce a generic platform for investigating fast driven rotating impurities in quantum many-body systems.
\end{abstract}

\maketitle


Superfluid helium droplets have emerged as powerful platforms for studying molecules embedded in a quantum environment~\cite{Grebenev1998, Grebenev2000, Nauta2001, Toennies2004, Schlesinger2010, Brauer2013, Pentlehner2013, Qiang2022}. Their weakly dissipative and highly coherent nature enables high-resolution spectroscopy and quantum control. They also allow fragile molecular species and complexes, relevant to chemistry, to be stabilized and probed under conditions inaccessible in the gas phase~\cite{Thaler2018, Albrechtsen2023, Choi2006}. Conversely, embedded molecules can serve as microscopic probes of the quantum many-body environment through their spectra and dynamics~\cite{Nancolas1985, Choi2006, Raston2012, Milner2023b}. They may also provide a route to characterizing the response of the surrounding helium when driven out of equilibrium, which remains an open challenge~\cite{Milner2023, Milner2023a}.

Moreover, these systems offer a controlled platform for studying impurities coupled to a quantum many-body environment. A paradigmatic example of impurity physics is the polaron, where a particle is dressed by collective excitations of the surrounding medium~\cite{Grusdt2016, Etrych2025, Massignan2026}. Molecules extend this paradigm because they possess internal rotational degrees of freedom and interact anisotropically with the bath. In superfluid helium, the coupling between molecular rotation and many-body excitations gives rise to the angulon quasiparticle: a rotor dressed by a cloud of superfluid excitations~\cite{Callegari1999, Lehmann2001, Szalewicz2008, Schmidt2015, Schmidt2016, Lemeshko2017}. At low angular momenta, this dressing is commonly described in terms of a renormalised rotational constant $B^*$ and an effective centrifugal distortion constant $D^*$. This equilibrium framework successfully captures low-energy rotational dynamics, including recent experiments on kicked molecules in superfluid helium~\cite{Chatterley2020, Cherepanov2021}.

In parallel, recent developments in driven polarons have shown that external forcing can be exploited to control quasiparticle properties and probe the surrounding environment~\cite{Vivanco2025, Mulkerin2025, Wasak2024, Hennebichler2026, Alhyder2026}. A molecular analogue could similarly exchange angular momentum with the surrounding superfluid, providing access to many-body dynamics beyond static spectral renormalisation. In this context, optical centrifuges provide a possible experimental platform: a chirped rotating electric field can trap the molecular axis and accelerate it adiabatically to very large angular momenta~\cite{Karczmarek1999, Larsen1999, Korobenko2014, Milner2017, MacPhail-Bartley2020, Wang2025a, MacPhail-Bartley2026}. Rotational frequencies far exceeding the bare rotational constant $B$, and potentially matching the characteristic bath dynamics, may therefore be experimentally achievable~\cite{MacPhail-Bartley2027}. This driven regime lies beyond the standard angulon description and remains largely unexplored theoretically for rotational degrees of freedom.

In this Letter, we tackle the problem of a strongly driven molecule embedded in a superfluid and investigate the interplay between molecular rotation and many-body bath dynamics. We show that rotation produces angular-momentum-resolved resonances arising from the rotational Doppler shift of bath excitations. These resonances generate a dissipative regime when the molecular rotation becomes comparable to the bath dynamics, and then act as a rotational sound barrier for the dressed molecule.

For concreteness, we consider a rotor trapped within a centrifuge rotating at constant frequency $\Omega$, as depicted in Fig.~1a. 
Before including the bath, we first investigate the driven rotor described in the co-rotating frame by
\beq \hat\H_{\rm rot}^{\rm RF} = B\hat{\tb{J}}^2 -\Omega \hat J_z -V_{\rm D}\sin^2(\hat\theta)\cos^2(\hat\phi) \eeq
where $\hat{\tb{J}}$ is the angular momentum operator of the rotor of spherical coordinates $(\theta, \phi)$ (see Fig.~1a). 
The first term corresponds to the free rotor and is characterised by angular momentum quantum numbers $(J,M)$ and energy spectrum $\E_J^0 = B J(J+1)$, which is $(2J+1)$-fold degenerate. 
The second term is the Coriolis contribution associated with the rotating frame, in which the Hamiltonian is time independent.
The last term corresponds to the drive, of amplitude $V_{\rm D}$, which aligns the molecule along the electric field at $\theta^*=\pi/2$ and $\phi^*=0, \pi$. 
Fig.~1a shows the resulting density distribution of the ground state of $\hat\H_{\rm rot}^{\rm RF} $. 
In the strongly trapped regime, the rotor co-rotates with the drive. 

Minimizing the energy over the dominant angular momentum components then yields $J^*\approx \Omega/2B$  (see SM Sec. 1).
In the lab frame, this corresponds to the rotational kinetic energy $\Omega^2/4B$, which must remain smaller than the trapping energy $V_{\rm D}$. 
This defines an upper bound to the rotational frequencies accessible with the centrifuge drive,
\beq \Omega  \lesssim \Omega_{\rm loc}=2\sqrt{B V_{\rm D}}, \eeq
 which can reach $10^2-10^3$~GHz in centrifuge experiments~\cite{Wang2025a}.
Below $\Omega_{\rm loc}$, the molecule remains polarised and follows the centrifuge, whereas above this scale the rotor progressively decouples from the drive.

Numerically, we diagonalize the full rotating-frame Hamiltonian for various rotation frequencies, $\hat\H_{\rm rot}^{\rm RF}|n(\Omega)\rangle=\E_n^{\rm RF}(\Omega)|n(\Omega)\rangle$, and estimate the corresponding  lab frame energy through $\E_n^{\rm LF} = \E_n^{\rm RF} +\Omega \langle n|\hat J_z|n\rangle$.
This allows us to identify the lab frame ground state as a function of the rotation frequency.
Fig.~1b shows its dominant rotational components, confirming that the molecule follows the drive up to $\Omega\sim \Omega_{\rm loc}$. 
Furthermore, the low-energy excitation spectrum, shown in Fig.~1c, reveals a characteristic spacing set by $\hbar  \Omega_{\rm loc}$.
Indeed, in the strongly trapped regime $\Omega \ll \Omega_{\rm loc}$, the molecule is strongly localized around $(\theta^*, \phi^*)$, and the Hamiltonian becomes locally harmonic in the small-angle limit, with harmonic frequency $\Omega_{\rm loc}$. 
Thus, for $\Omega \lesssim\Omega_{\rm loc}$ the low-energy spectrum corresponds to an angularly localized pendular state~\cite{Rost1992, Kim1996, Redchenko2016}, while at larger rotation frequencies the rotor delocalizes.

We now couple the driven rotor to a bosonic bath of dispersion $\w_k$.
To access the regime where molecular rotation becomes comparable to the bath dynamics, we consider strong drives such that $\Omega_{\rm loc}$ exceeds the characteristic excitation energies of the bath.
Setting $\hbar=1$, the total Hamiltonian in the co-rotating frame reads~\cite{Schmidt2015}:
\beqa \hat\H^{\rm RF} &=& \hat\H_{\rm rot}^{\rm RF}+   \sum_{k\lambda\mu}
    (\omega_k-\Omega\mu)
    \hat b^\dagger_{k\lambda\mu}
    \hat b_{k\lambda\mu} \ldots\\
    &&  \ldots + \sum_{k\lambda\mu}
  U_\lambda(k)
    \left[
        Y_{\lambda\mu}(\hat\theta,\hat\phi)
        \hat b_{k\lambda\mu}
        +
        Y^\dagger_{\lambda\mu}(\hat\theta,\hat\phi)
        \hat b^\dagger_{k\lambda\mu}
    \right]\nonumber
\eeqa
where $ b^\dagger_{k\lambda\mu}$ is the creation operation of a boson of linear momentum $k$ and angular momentum $\lambda$ with projection $\mu$ onto the $z$-axis, $ Y_{\lambda\mu}$ is the spherical harmonic and $U_\lambda(k)$ is the spherical Bessel-transformed molecule-bath interaction potential (see SM Sec. 2). 
In practice, the molecule--bath interaction is often dominated by the quadrupolar channel $\lambda=2$~\cite{Lemeshko2017}.
Fig.~2d shows a typical profile of $U_2(k)$ together with the dispersion of helium~\cite{Donnelly1981}.  
For simplicity, we consider only this channel and focus on symmetric molecules, for which only even rotational states contribute.

\begin{figure}
	\centering
	\includegraphics{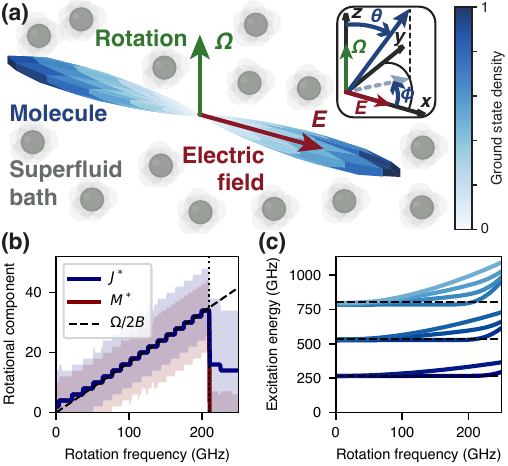}
	\caption{ \tb{System and effect of the drive.}
	\tb{(a)} Schematics of the system: a molecule embedded in a superfluid bath is polarised by an electric field $E$ rotating at frequency $\Omega$ in the xy-plane. 
	In blue is the molecular ground state density $\hat\H_{\rm rot}^{\rm RF}$ at $\Omega=0$ for $B=3$ GHz and $V_{\rm D}=6$ THz. 
	(Inset) Detail of the geometry and definition of the spherical coordinates $(\theta, \phi)$. In the co-rotating frame the electric field is along the $x$-direction while its rotation vector still is in the $z$-direction.
	\tb{(b)} Dominant angular momentum component $J^*$ and its projection onto the $z$-axis $M^*$ of the ground state of $\hat\H_{\rm rot}^{\rm RF}$ as a function of the rotation frequency $\Omega$. 
	The shaded region contains 95\% of the ground-state weight. 
	The fall happens at $\Omega\approx 0.8\,\Omega_{\rm loc}$. 
	\tb{(c)} Low-energy excitation spectrum $\E_n^{\rm RF}-\E_0^{\rm RF}$ as a function of the rotation frequency $\Omega$. Dashed lines indicate the harmonic level spacing $\Omega_{\rm loc}$. 
	 	}\label{fig1}
\end{figure}

In conventional angulon theory, weak-coupling approaches typically apply to molecules embedded in Bose--Einstein condensates, whereas molecules in superfluid helium often require strong-coupling treatments~\cite{Schmidt2015,Schmidt2016,Lemeshko2017}.
The weak-coupling approach is valid when the characteristic energy level spacing $\Delta \E$ of the bare rotor exceeds the interaction energy scale $\bar U$~\cite{Lemeshko2017}. 
For free molecules, $\Delta \E \sim B$, which typically ranges from 1 to $10^2$~GHz, while molecule--helium interactions are commonly of order $\bar U \sim 10^2$~GHz. 
In the driven regime, however, the strong confinement imposed by the centrifuge increases the characteristic level spacing to $\Delta \E \sim \Omega_{\rm loc}$, which typically exceeds  $\bar U$~\cite{Wang2025a}.
Consequently, despite the strong molecule--helium interaction, the bath action can be treated perturbatively over a broad range of experimentally relevant parameters when the rotor is strongly driven.

Adapting the perturbative approach developed for the angulon~\cite{Bighin2017} to the driven rotor, the retarded Green's function in the co-rotating frame can be written as  (see SM Sec. 3):
\beq \hat G^{\rm RF}(\w)=\frac{1}{\w - \hat\H_{\rm rot}^{\rm RF}-\hat \Sigma(\w) + i0^+}\eeq
where $\hat \Sigma(\w)$ is the self-energy operator. To second order in the molecule--bath interaction, its matrix elements in the eigenbasis of $\hat\H_{\rm rot}^{\rm RF}$ read
\beq     \langle m|\hat\Sigma(\omega) |n\rangle
    =
    \sum_{k,\lambda,\mu, l}
    \frac{ |U_\lambda(k)|^2
        \langle m|
        \hat Y_{\lambda\mu}
        |l\rangle   \langle l|
        \hat Y_{\lambda\mu}^{\dagger}
        |n\rangle
    }{
        \omega
        -\omega_k
        +\Omega\mu
        -\E_l^{\rm RF}
        +i0^+
    } 
    \label{selfenergy}
    \eeq
Here, we assumed zero temperature, although finite-temperature generalizations are straightforward~\cite{Rammer2007}. 
The self-energy describes virtual transitions of the driven rotor mediated by the emission and reabsorption of bath excitations carrying angular momentum $(\lambda,\mu)$. 
Importantly, the rotating frame shifts the bath excitation energies by $\Omega\mu$, leading to angular-momentum-resolved resonances absent in equilibrium angulon theory.
Note also that both the eigenvectors and eigenvalues of $\hat\H_{\rm rot}^{\rm RF}$ depend on the rotation frequency. 

Using the dispersion relation of superfluid helium (Fig.~2d)~\cite{Donnelly1981}, we compute the spectral function of the rotor in the co-rotating frame,
$A(\omega)=-\frac{1}{\pi}\textnormal{Tr}[\hat G^{\rm RF}(\omega)]$
as a function of the frequency detuning
$\Delta\omega=\omega-\E_0^{\rm RF}(\Omega)$, 
as shown in Fig.~2a.
Focusing on the ground-state branch, we identify four distinct regimes (separated by the vertical dashed lines in Fig.~2a). At low rotation frequencies, the bath dynamics remains fast compared with the molecular rotation. The trapped rotor is therefore only weakly renormalised by the superfluid, forming a pendulon state~\cite{Rost1992, Kim1996, Redchenko2016}. 
This regime was recently explored experimentally for CS$_2$ and OCS molecules embedded in superfluid helium and accelerated by an optical centrifuge to frequencies of up to 40~GHz~\cite{MacPhail-Bartley2027}. The measurements were rationalised by attributing to the bath a slow thermalisation of the driven rotor towards its polarised ground state, consistent with our model.
At intermediate frequencies, the linewidth broadens substantially, as further illustrated in Fig.~2b. This signals a short quasiparticle lifetime and dissipation into the bath. For the helium parameters considered here, this dissipative regime extends approximately from 60 to 135~GHz. Although the precise location of its boundaries depends on the molecule--bath interaction potential, the existence of this regime does not. Beyond this window, we observe a transition towards a new stable quasiparticle. Finally, as the rotation frequency approaches $\Omega_{\rm loc}$, the rotor progressively delocalizes and the spectral branches become periodic in the co-rotating frame. This indicates that quasiparticle excitations are no longer well defined in the rotating frame, but instead recover a simpler interpretation in the lab frame.

\begin{figure}
	\centering
	\includegraphics{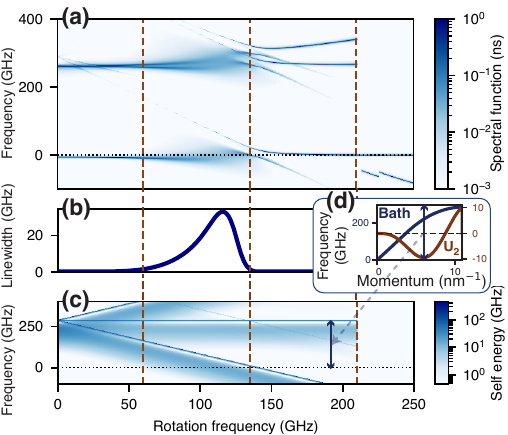}
	\caption{ \tb{Many-body renormalisation of the driven rotor by superfluid helium.}
	\tb{(a)} Full spectral function $A(\w)$ as a function of the ground state frequency detuning $\Delta\w=\w-\E_0^{\rm RF}(\Omega)$ and the rotation frequency $\Omega$. We used $B=3$ GHz and $V_{\rm D}=6$ THz. 
	\tb{(b)} Linewidth of the ground state extracted from the spectral function as a function of the rotation frequency $\Omega$. 
	\tb{(c)} Imaginary part of the ground state self energy $-\tn{Im}\left[\Sigma_0(\w)\right]$ as a function of the frequency detuning $\Delta\w=\w-\E_0^{\rm RF}(\Omega)$ and the rotation frequency $\Omega$.  
\tb{(a-c)} Vertical dashed lines at $\Omega=60$, 135, and 210~GHz are guides to the eye indicating the successive dynamical regimes discussed in the text. 
\tb{(d)} Dispersion $\w_k$ of superfluid helium together with the spherical Bessel-transformed molecule-bath van der Waals interaction potential $U_2(k)$ as a function of the momentum $k$. The dashed blue lines are guides for the eye of the impact of the bath dispersion on the self energy.
	}\label{fig2}
\end{figure}

To rationalise this phenomenology, we analyse the structure of the self-energy.
In the weak-coupling regime, the rotor eigenmodes are only weakly renormalised by the bath and the self-energy can therefore be analyzed mode by mode through the diagonal projection
$\Sigma_n(\omega)=\langle n|\hat\Sigma(\omega)|n\rangle.$
The self-energy mainly contributes on shell, namely for small frequency detuning
$\Delta\omega=\omega-\E_n^{\rm RF}.$
In the strongly trapped regime, the neighbouring excited rotational states are separated by the harmonic scale $\Omega_{\rm loc}$, which is too large to be resonantly accessed. 
As a result, the dominant contribution arises from virtual transitions through the state itself ($l=n$ in Eq.~\eqref{selfenergy}), leading to the resonance condition
\beq
\Delta\omega
=
\omega_k-\Omega\mu.
\eeq
Fig.~2c shows the imaginary part of the ground-state self-energy for a molecule embedded in superfluid helium. 
At $\Omega=0$, the self-energy is significant only for frequency detunings comparable to the bath excitation energies (Fig.~2d). 
At finite rotation frequencies, however, the bath excitations become Doppler shifted, generating several resonance branches associated with the angular momentum projections $\mu=-2,0,2$. 
When one of these branches crosses zero, exciting the bath no longer requires any quasienergy, leading to a strong enhancement of boson emission. 
However, these bosons still carry positive energy in the lab frame and therefore provide a dissipation channel.
Consequently, the rotor enters a regime of large dissipation and strong coupling to the bath. 
This is the rotational analogue of a sound barrier: the molecule rotates faster than the characteristic propagation velocity of excitations supported by the surrounding medium. 
Since the bath dispersion is continuous, however, this transition is broadened over the range of rotational frequencies $\Omega_{\rm diss}$ satisfying the self-consistent equation
\beq
\Omega_{\rm diss}=\frac{\omega_k-\Delta \E_0^{\rm RF}(\Omega_{\rm diss})}{2} 
\label{w_diss}
\eeq
for the wavevectors $k$ significantly coupled to the molecule (see Fig.~2d). 
Here, $\Delta \E_0^{\rm RF}$ denotes the bath-induced energy shift of the ground state arising from the real part of the self-energy. 
For superfluid helium, this correction remains small compared to the characteristic bath excitation energies.
Finally, beyond the dissipative regime, the Doppler-shifted angular-momentum channels are pushed away from resonance.
The remaining contribution is dominated by the $\mu=0$ channel, corresponding to bath excitations carrying no angular momentum along the rotation axis and therefore unaffected by the Doppler shift, resulting in a modified pendulon state.

While the previous results were obtained for the dispersion relation of superfluid helium, the same phenomenology also emerges for a Bose--Einstein condensate (BEC), albeit on a different frequency scale. Indeed, the Bogoliubov dispersion $\omega_k=\sqrt{\epsilon_k(\epsilon_k + 8\pi n a/m)}$ where $\epsilon_k=k^2/2m$ is the kinetic energy, $n$ is the atom density of mass $m$, and $a$ the scattering length, is typically several orders of magnitude smaller than in helium. 
To connect these two limits, we consider densities intermediate between typical BECs and superfluid helium within the Bogoliubov description.
A representative spectral function for a driven molecule in a BEC is shown in Fig.~3a, with the corresponding linewidths in Fig.~3b.
Focusing on the ground state, the same dynamical regimes remain clearly visible, as indicated by the dashed lines. This demonstrates that the rotational sound barrier is a generic consequence of the competition between molecular rotation and the propagation of bath excitations.
We now turn to the first excited state near $\Omega_{\rm loc}$, which forms a nearly degenerate doublet. While one branch follows the same dissipative window as the ground state, the second exhibits an additional dissipative resonance at higher rotation frequencies, around $\Omega\approx7$~GHz. This feature originates from the $\mu=-2$ Doppler-shifted bath branch, which enables resonant transitions between the ground and excited states when
\beq
\Omega_{\rm diss}\approx \frac{\E_1^{\rm RF}-\E_0^{\rm RF}-\omega_k+\Delta \E_0^{\rm RF}(\Omega_{\rm diss})}{2}.
\eeq
Molecular rotation therefore not only modifies the ground-state dressing, but also opens additional dissipative channels involving excited rotational states.

\begin{figure}
	\centering
	\includegraphics{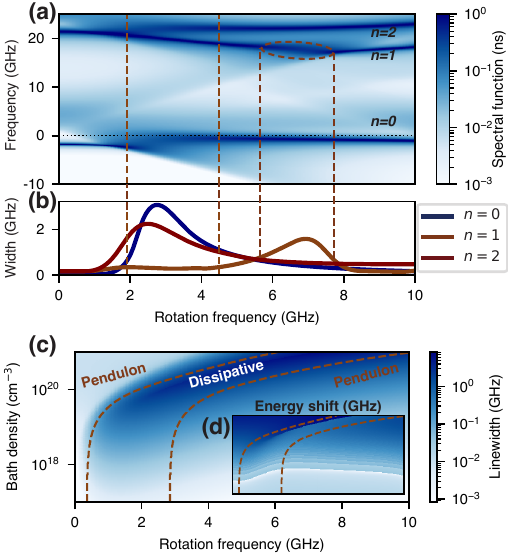}
	\caption{ \tb{Dynamical regimes of the driven rotor embedded in a BEC.}
	\tb{(a)} Full spectral function $A(\w)$ as a function of the ground state frequency detuning $\Delta\w=\w-\E_0^{\rm RF}(\Omega)$ and the rotation frequency $\Omega$, at density density $n=10^{20}~\mathrm{cm}^{-3}$. We used $B=3$ GHz and $V_{\rm D}=50$ GHz and a BEC with the mass of Li, $m=7\, m_0$ ($m_0$ being the atomic mass) and a scattering length of $a = 10 \, a_0$ ($a_0$ being the Bohr radius).
	\tb{(b)} Linewidth of the ground state and the two first excited states extracted from the spectral function as a function of the rotation frequency $\Omega$. 
\tb{(c)} Linewidth of the ground state extracted from the spectral function as a function of the superfluid density $n$ and the rotation frequency $\Omega$. 
\tb{(d)} Energy shift $-\Delta \E_0^{\rm RF}$ of the ground state due to the bath as a function of the superfluid density $n$ and the rotation frequency $\Omega$. Same axis and colorbar than in (c). 
\tb{(a-d)} Dashed lines are guides to the eyes indicating the boundaries of the different dynamical regimes. Their positions follow from the resonance condition $\Omega=\omega_k/2-\Delta \E_0^{\rm RF}/2$, evaluated for $k=1$ and $2.8$~nm$^{-1}$ and the peak value of $\Delta \E_0^{\rm RF}(n)$. 
	}\label{fig3}
\end{figure}

Finally, by sweeping the density, we construct a dynamical diagram for the driven molecule in a superfluid. Figs.~3c-d show the quasiparticle linewidth and bath-induced energy shift of the ground state. The discrete steps in the energy shift originate from the discrete rotational spectrum of the molecule, whose level spacing remains comparable to the other energy scales at low densities. From these plots, we identify approximate boundaries of the dissipative regime, shown by dashed lines. These boundaries are quantitatively governed by Eq.~\eqref{w_diss}.
At low densities, the bath-induced energy shift remains small and the resonance condition is dominated by the bath dispersion, whose frequency increases with density.
 At larger densities, however, the bath-induced energy shift becomes comparable to the excitation energy and contributes significantly to the resonance condition. 
 Because $\Delta \E_0^{\rm RF}$ decreases approximately linearly with density, the barrier frequency then increases more rapidly than expected from the Bogoliubov dispersion alone.

To summarize, we investigated a driven molecule rotating in a superfluid environment. Within the experimentally relevant platform of an optical centrifuge, we identified a characteristic harmonic frequency that can exceed the bath excitation energies, enabling molecular rotation to become resonant with superfluid modes. These resonances give rise to a rotational sound barrier, associated with enhanced angular-momentum exchange and energy dissipation into the bath. 
We further demonstrated that this mechanism is generic, persisting from superfluid helium to Bose--Einstein condensates. 
Recent optical-centrifuge experiments on molecules embedded in superfluid helium offer a promising route to accessing this regime and testing the predicted resonant dissipation~\cite{MacPhail-Bartley2027}.
 Taken together, our results establish a framework for driven rotational impurities in quantum many-body environments and open the door to non-perturbative numerical approaches~\cite{Bighin2018}. More broadly, our results suggest that ultrafast nonequilibrium impurity dynamics can be used to probe the dynamics of quantum solvents beyond equilibrium, including bath dispersion, Landau-type critical behaviour, angular-momentum and energy flow, and quasiparticle dynamics.


\section*{Acknowledgements} 

The authors thank Valery Milner, Henrik Stapelfeldt, Volker Karle, Jing-Lun Li and Georgios M. Koutentakis for fruitful discussions. 
B.C. acknowledges support from the NOMIS Foundation. R.\ A. acknowledges funding from the Austrian Academy of Science \"{O}AW grant No. PR1029OEAW03.
GB acknowledges support from the project “Implementation of cutting-edge research and its application as part of the Scientific Center of Excellence for Quantum and Complex Systems, and Representations of Lie Algebras“, Grant No. PK.1.1.10.0004, co-financed by the European Union through the European Regional Development Fund - Competitiveness and Cohesion Programme 2021-2027.

\bibliography{bibfile}

@article{Albrechtsen2023,
  title = {Observing the Primary Steps of Ion Solvation in Helium Droplets},
  author = {Albrechtsen, Simon H. and Schouder, Constant A. and Vi{\~n}as Mu{\~n}oz, Alberto and Christensen, Jeppe K. and Engelbrecht Petersen, Christian and Pi, Mart{\'i} and Barranco, Manuel and Stapelfeldt, Henrik},
  year = 2023,
  month = nov,
  journal = {Nature},
  volume = {623},
  number = {7986},
  pages = {319--323},
  publisher = {Nature Publishing Group},
  issn = {1476-4687},
  doi = {10.1038/s41586-023-06593-5},
  urldate = {2024-05-28},
  copyright = {2023 The Author(s), under exclusive licence to Springer Nature Limited},
  langid = {english}
}

@article{Alhyder2026,
  title = {Phenomenological Model of Decaying {{Bose}} Polarons},
  author = {Alhyder, R. and Bruun, G. M. and Pohl, T. and Lemeshko, M. and Volosniev, A. G.},
  year = 2026,
  month = feb,
  journal = {Physical Review Research},
  volume = {8},
  number = {1},
  pages = {L012034},
  publisher = {American Physical Society},
  doi = {10.1103/16dk-5dgx},
  urldate = {2026-08-26}
}

@article{Bighin2017,
  title = {Diagrammatic Approach to Orbital Quantum Impurities Interacting with a Many-Particle Environment},
  author = {Bighin, G. and Lemeshko, M.},
  year = 2017,
  month = aug,
  journal = {Physical Review B},
  volume = {96},
  number = {8},
  pages = {085410},
  publisher = {American Physical Society},
  doi = {10.1103/PhysRevB.96.085410},
  urldate = {2026-02-09}
}

@article{Bighin2018,
  title = {Diagrammatic {{Monte Carlo Approach}} to {{Angular Momentum}} in {{Quantum Many-Particle Systems}}},
  author = {Bighin, G. and Tscherbul, T. V. and Lemeshko, M.},
  year = 2018,
  month = oct,
  journal = {Physical Review Letters},
  volume = {121},
  number = {16},
  pages = {165301},
  publisher = {American Physical Society},
  doi = {10.1103/PhysRevLett.121.165301},
  urldate = {2026-02-09}
}

@article{Brauer2013,
  title = {Critical {{Landau Velocity}} in {{Helium Nanodroplets}}},
  author = {Brauer, Nils B. and Smolarek, Szymon and Loginov, Evgeniy and Mateo, David and Hernando, Alberto and Pi, Marti and Barranco, Manuel and Buma, Wybren J. and Drabbels, Marcel},
  year = 2013,
  month = oct,
  journal = {Physical Review Letters},
  volume = {111},
  number = {15},
  pages = {153002},
  publisher = {American Physical Society},
  doi = {10.1103/PhysRevLett.111.153002},
  urldate = {2026-06-08}
}

@article{Callegari1999,
  title = {Superfluid {{Hydrodynamic Model}} for the {{Enhanced Moments}} of {{Inertia}} of {{Molecules}} in {{Liquid}} \$\textbraceleft\textbraceright\textasciicircum\textbraceleft 4\textbraceright\textbackslash mathrm\textbraceleft{{He}}\textbraceright\$},
  author = {Callegari, C. and Conjusteau, A. and Reinhard, I. and Lehmann, K. K. and Scoles, G. and Dalfovo, F.},
  year = 1999,
  month = dec,
  journal = {Physical Review Letters},
  volume = {83},
  number = {24},
  pages = {5058--5061},
  publisher = {American Physical Society},
  doi = {10.1103/PhysRevLett.83.5058},
  urldate = {2026-06-08}
}

@article{Chatterley2020,
  title = {Rotational {{Coherence Spectroscopy}} of {{Molecules}} in {{Helium Nanodroplets}}: {{Reconciling}} the {{Time}} and the {{Frequency Domains}}},
  shorttitle = {Rotational {{Coherence Spectroscopy}} of {{Molecules}} in {{Helium Nanodroplets}}},
  author = {Chatterley, Adam S. and Christiansen, Lars and Schouder, Constant A. and J{\o}rgensen, Anders V. and Shepperson, Benjamin and Cherepanov, Igor N. and Bighin, Giacomo and Zillich, Robert E. and Lemeshko, Mikhail and Stapelfeldt, Henrik},
  year = 2020,
  month = jul,
  journal = {Physical Review Letters},
  volume = {125},
  number = {1},
  pages = {013001},
  publisher = {American Physical Society},
  doi = {10.1103/PhysRevLett.125.013001},
  urldate = {2026-02-24}
}

@article{Cherepanov2021,
  title = {Excited Rotational States of Molecules in a Superfluid},
  author = {Cherepanov, Igor N. and Bighin, Giacomo and Schouder, Constant A. and Chatterley, Adam S. and Albrechtsen, Simon H. and Mu{\~n}oz, Alberto Vi{\~n}as and Christiansen, Lars and Stapelfeldt, Henrik and Lemeshko, Mikhail},
  year = 2021,
  month = dec,
  journal = {Physical Review A},
  volume = {104},
  number = {6},
  pages = {L061303},
  publisher = {American Physical Society},
  doi = {10.1103/PhysRevA.104.L061303},
  urldate = {2024-05-28}
}

@article{Choi2006,
  title = {Infrared Spectroscopy of Helium Nanodroplets: Novel Methods for Physics and Chemistry},
  shorttitle = {Infrared Spectroscopy of Helium Nanodroplets},
  author = {Choi, M. Y. and Douberly, G. E. and Falconer, T. M. and Lewis, W. K. and Lindsay, C. M. and Merritt, J. M. and Stiles, P. L. and Miller, R. E.},
  year = 2006,
  month = jan,
  journal = {International Reviews in Physical Chemistry},
  volume = {25},
  number = {1-2},
  pages = {15--75},
  publisher = {Taylor \& Francis},
  issn = {0144-235X},
  doi = {10.1080/01442350600625092},
  urldate = {2024-05-16}
}

@article{Donnelly1981,
  title = {Specific Heat and Dispersion Curve for Helium {{II}}},
  author = {Donnelly, R. J. and Donnelly, J. A. and Hills, R. N.},
  year = 1981,
  month = sep,
  journal = {Journal of Low Temperature Physics},
  volume = {44},
  number = {5},
  pages = {471--489},
  issn = {1573-7357},
  doi = {10.1007/BF00117839},
  urldate = {2025-02-24},
  langid = {english}
}

@article{Etrych2025,
  title = {Universal {{Quantum Dynamics}} of {{Bose Polarons}}},
  author = {Etrych, Ji{\v r}{\'i} and Martirosyan, Gevorg and Cao, Alec and Ho, Christopher J. and Hadzibabic, Zoran and Eigen, Christoph},
  year = 2025,
  month = may,
  journal = {Physical Review X},
  volume = {15},
  number = {2},
  pages = {021070},
  publisher = {American Physical Society},
  doi = {10.1103/PhysRevX.15.021070},
  urldate = {2026-06-16}
}

@article{Farrokhpour2013,
  title = {Ab Initio Intermolecular Potential Energy Surfaces of {{He}}--{{CS2}}, {{Ne}}--{{CS2}} and {{Ar}}--{{CS2}} Complexes},
  author = {Farrokhpour, H. and Tozihi, M.},
  year = 2013,
  month = mar,
  journal = {Molecular Physics},
  volume = {111},
  number = {6},
  pages = {779--791},
  publisher = {Taylor \& Francis},
  issn = {0026-8976},
  doi = {10.1080/00268976.2012.745630},
  urldate = {2026-03-20}
}

@article{Grebenev1998,
  title = {Superfluidity {{Within}} a {{Small Helium-4 Cluster}}: {{The Microscopic Andronikashvili Experiment}}},
  shorttitle = {Superfluidity {{Within}} a {{Small Helium-4 Cluster}}},
  author = {Grebenev, Slava and Toennies, J. Peter and Vilesov, Andrei F.},
  year = 1998,
  month = mar,
  journal = {Science},
  volume = {279},
  number = {5359},
  pages = {2083--2086},
  publisher = {American Association for the Advancement of Science},
  doi = {10.1126/science.279.5359.2083},
  urldate = {2026-06-08}
}

@article{Grebenev2000,
  title = {The Rotational Spectrum of Single {{OCS}} Molecules in Liquid {{4He}} Droplets},
  author = {Grebenev, Slava and Hartmann, Matthias and Havenith, Martina and Sartakov, Boris and Toennies, J. Peter and Vilesov, Andrei F.},
  year = 2000,
  month = mar,
  journal = {The Journal of Chemical Physics},
  volume = {112},
  number = {10},
  pages = {4485--4495},
  issn = {0021-9606},
  doi = {10.1063/1.481011},
  urldate = {2026-05-26}
}

@incollection{Grusdt2016,
  title = {New Theoretical Approaches to {{Bose}} Polarons},
  booktitle = {Quantum {{Matter}} at {{Ultralow Temperatures}}},
  author = {Grusdt, F. and Demler, E.},
  year = 2016,
  pages = {325--411},
  publisher = {IOS Press},
  doi = {10.3254/978-1-61499-694-1-325},
  urldate = {2026-05-29},
  langid = {english}
}

@misc{Hennebichler2026,
  title = {The Moving {{Fermi}} Polaron},
  author = {Hennebichler, Johanna and Erlenstedt, Ruben and Dobler, Erich and Baroni, Cosetta and Grimm, Rudolf and Caldara, Matteo and Bruun, Georg and Massignan, Pietro},
  year = 2026,
  month = jun,
  number = {arXiv:2606.21567},
  eprint = {2606.21567},
  primaryclass = {cond-mat.quant-gas},
  publisher = {arXiv},
  doi = {10.48550/arXiv.2606.21567},
  urldate = {2026-06-23},
  archiveprefix = {arXiv}
}

@article{Karczmarek1999,
  title = {Optical {{Centrifuge}} for {{Molecules}}},
  author = {Karczmarek, Joanna and Wright, James and Corkum, Paul and Ivanov, Misha},
  year = 1999,
  month = apr,
  journal = {Physical Review Letters},
  volume = {82},
  number = {17},
  pages = {3420--3423},
  publisher = {American Physical Society},
  doi = {10.1103/PhysRevLett.82.3420},
  urldate = {2026-06-08}
}

@article{Kim1996,
  title = {Spectroscopy of Pendular States in Optical-field-aligned Species},
  author = {Kim, Wousik and Felker, Peter M.},
  year = 1996,
  month = jan,
  journal = {The Journal of Chemical Physics},
  volume = {104},
  number = {3},
  pages = {1147--1150},
  issn = {0021-9606},
  doi = {10.1063/1.470770},
  urldate = {2026-06-16}
}

@article{Korobenko2014,
  title = {Rotational Spectroscopy with an Optical Centrifuge},
  author = {Korobenko, Aleksey and A.~Milner, Alexander and W.~Hepburn, John and Milner, Valery},
  year = 2014,
  journal = {Physical Chemistry Chemical Physics},
  volume = {16},
  number = {9},
  pages = {4071--4076},
  publisher = {Royal Society of Chemistry},
  doi = {10.1039/C3CP54598A},
  urldate = {2026-05-26},
  langid = {english}
}

@article{Larsen1999,
  title = {Aligning Molecules with Intense Nonresonant Laser Fields},
  author = {Larsen, Jakob Juul and Sakai, Hirofumi and Safvan, C. P. and {Wendt-Larsen}, Ida and Stapelfeldt, Henrik},
  year = 1999,
  month = nov,
  journal = {The Journal of Chemical Physics},
  volume = {111},
  number = {17},
  pages = {7774--7781},
  issn = {0021-9606},
  doi = {10.1063/1.480112},
  urldate = {2026-06-08}
}

@article{Lehmann2001,
  title = {Rotation in Liquid {{He4 Lessons}} from a Highly Simplified Model},
  author = {Lehmann, Kevin K.},
  year = 2001,
  month = mar,
  journal = {The Journal of Chemical Physics},
  volume = {114},
  number = {10},
  pages = {4643--4648},
  issn = {0021-9606},
  doi = {10.1063/1.1334620},
  urldate = {2026-06-08}
}

@article{Lemeshko2017,
  title = {Quasiparticle {{Approach}} to {{Molecules Interacting}} with {{Quantum Solvents}}},
  author = {Lemeshko, Mikhail},
  year = 2017,
  month = feb,
  journal = {Physical Review Letters},
  volume = {118},
  number = {9},
  pages = {095301},
  publisher = {American Physical Society},
  doi = {10.1103/PhysRevLett.118.095301},
  urldate = {2024-02-26}
}

@article{MacPhail-Bartley2020,
  title = {Laser Control of Molecular Rotation: {{Expanding}} the Utility of an Optical Centrifuge},
  shorttitle = {Laser Control of Molecular Rotation},
  author = {{MacPhail-Bartley}, Ian and Wasserman, Walter W. and Milner, Alexander A. and Milner, Valery},
  year = 2020,
  month = apr,
  journal = {Review of Scientific Instruments},
  volume = {91},
  number = {4},
  pages = {045122},
  issn = {0034-6748},
  doi = {10.1063/1.5140358},
  urldate = {2026-06-08}
}

@article{MacPhail-Bartley2026,
  title = {Control of {{Molecular Rotation}} in {{Helium Nanodroplets}} with an {{Optical Centrifuge}}},
  author = {{MacPhail-Bartley}, Ian and Milner, Alexander A. and Stienkemeier, Frank and Milner, Valery},
  year = 2026,
  month = jan,
  journal = {Physical Review Letters},
  volume = {136},
  number = {3},
  pages = {033002},
  publisher = {American Physical Society},
  doi = {10.1103/5jnj-97vs},
  urldate = {2026-03-16}
}

@misc{MacPhail-Bartley2027,
  title = {Optical Centrifuge as a Probe of Strong Dissipative Coupling between a Molecular Rotor and Superfluid Helium},
  author = {{MacPhail-Bartley}, Ian and Mahr, S{\"o}ren E. and Peters, Cameron E. and Coquinot, Baptiste and Karle, Volker and Bighin, Giacomo and Alhyder, Ragheed and Lemeshko, Mikhail and Stapelfeldt, Henrik and Milner, Valery},
  year = 2026,
  month = aug,
  number = {arXiv:2608.25161},
  eprint = {2608.25161},
  primaryclass = {quant-ph},
  publisher = {arXiv},
  doi = {10.48550/arXiv.2608.25161},
  urldate = {2026-08-27},
  archiveprefix = {arXiv}
}

@article{Massignan2026,
  title = {Polarons in Atomic Gases and Two-Dimensional Semiconductors},
  author = {Massignan, Pietro and Schmidt, Richard and Astrakharchik, Grigori E. and {\.I}mamoglu, Ata{\c c} and Zwierlein, Martin and Arlt, Jan J. and Bruun, Georg M.},
  year = 2026,
  month = apr,
  journal = {Reviews of Modern Physics},
  publisher = {American Physical Society},
  doi = {10.1103/4nng-bb9z}
}

@article{Milner2017,
  title = {Probing Molecular Potentials with an Optical Centrifuge},
  author = {Milner, A. A. and Korobenko, A. and Hepburn, J. W. and Milner, V.},
  year = 2017,
  month = sep,
  journal = {The Journal of Chemical Physics},
  volume = {147},
  number = {12},
  pages = {124202},
  issn = {0021-9606},
  doi = {10.1063/1.5004788},
  urldate = {2026-06-08}
}

@article{Milner2023,
  title = {Ultrafast Nonequilibrium Dynamics of Rotons in Superfluid Helium},
  author = {Milner, Alexander A. and Stamp, Philip C. E. and Milner, Valery},
  year = 2023,
  month = apr,
  journal = {Proceedings of the National Academy of Sciences},
  volume = {120},
  number = {17},
  pages = {e2303231120},
  publisher = {Proceedings of the National Academy of Sciences},
  doi = {10.1073/pnas.2303231120},
  urldate = {2024-04-16}
}

@article{Milner2023a,
  title = {Controlled {{Excitation}} of {{Rotons}} in {{Superfluid Helium}} with an {{Optical Centrifuge}}},
  author = {Milner, Alexander A. and Milner, Valery},
  year = 2023,
  month = oct,
  journal = {Physical Review Letters},
  volume = {131},
  number = {16},
  pages = {166001},
  publisher = {American Physical Society},
  doi = {10.1103/PhysRevLett.131.166001},
  urldate = {2024-04-16}
}

@article{Milner2023b,
  title = {Dynamics of Molecular Rotors in Bulk Superfluid Helium},
  author = {Milner, Alexander A. and Apkarian, V. A. and Milner, Valery},
  year = 2023,
  month = jun,
  journal = {Science Advances},
  volume = {9},
  number = {26},
  pages = {eadi2455},
  publisher = {American Association for the Advancement of Science},
  doi = {10.1126/sciadv.adi2455},
  urldate = {2026-06-08}
}

@article{Mulkerin2025,
  title = {Hybrid-Pair Superfluidity in a Strongly Driven {{Fermi}} Gas},
  author = {Mulkerin, Brendan C. and Bleu, Olivier and Cabrera, Cesar R. and Parish, Meera M. and Levinsen, Jesper},
  year = 2025,
  month = jul,
  journal = {Physical Review A},
  volume = {112},
  number = {1},
  pages = {013314},
  publisher = {American Physical Society},
  doi = {10.1103/xtfy-8nrt},
  urldate = {2026-06-16}
}

@article{Nancolas1985,
  title = {A New Form of Energy Dissipation by a Moving Object in {{He II}}},
  author = {Nancolas, G. G. and Ellis, T. and McClintock, P. V. E. and Bowley, R. M.},
  year = 1985,
  month = aug,
  journal = {Nature},
  volume = {316},
  number = {6031},
  pages = {797--799},
  publisher = {Nature Publishing Group},
  issn = {1476-4687},
  doi = {10.1038/316797a0},
  urldate = {2026-06-08},
  copyright = {1985 Springer Nature Limited},
  langid = {english}
}

@article{Nauta2001,
  title = {Rotational and Vibrational Dynamics of {{CO2}} and {{N2O}} in Helium Nanodroplets},
  author = {Nauta, K. and Miller, R. E.},
  year = 2001,
  month = dec,
  journal = {The Journal of Chemical Physics},
  volume = {115},
  number = {22},
  pages = {10254--10260},
  issn = {0021-9606},
  doi = {10.1063/1.1416492},
  urldate = {2026-05-26}
}

@article{Pentlehner2013,
  title = {Impulsive {{Laser Induced Alignment}} of {{Molecules Dissolved}} in {{Helium Nanodroplets}}},
  author = {Pentlehner, Dominik and Nielsen, Jens H. and Slenczka, Alkwin and M{\o}lmer, Klaus and Stapelfeldt, Henrik},
  year = 2013,
  month = mar,
  journal = {Physical Review Letters},
  volume = {110},
  number = {9},
  pages = {093002},
  publisher = {American Physical Society},
  doi = {10.1103/PhysRevLett.110.093002},
  urldate = {2026-06-08}
}

@article{Qiang2022,
  title = {Femtosecond {{Rotational Dynamics}} of \$\textbraceleft\textbackslash mathrm\textbraceleft{{D}}\textbraceright\textbraceright\_\textbraceleft 2\textbraceright\$ {{Molecules}} in {{Superfluid Helium Nanodroplets}}},
  author = {Qiang, Junjie and Zhou, Lianrong and Lu, Peifen and Lin, Kang and Ma, Yongzhe and Pan, Shengzhe and Lu, Chenxu and Jiang, Wenyu and Sun, Fenghao and Zhang, Wenbin and Li, Hui and Gong, Xiaochun and Averbukh, Ilya Sh. and Prior, Yehiam and Schouder, Constant A. and Stapelfeldt, Henrik and Cherepanov, Igor N. and Lemeshko, Mikhail and J{\"a}ger, Wolfgang and Wu, Jian},
  year = 2022,
  month = jun,
  journal = {Physical Review Letters},
  volume = {128},
  number = {24},
  pages = {243201},
  publisher = {American Physical Society},
  doi = {10.1103/PhysRevLett.128.243201},
  urldate = {2026-06-08}
}

@book{Rammer2007,
  title = {Quantum {{Field Theory}} of {{Non-equilibirum States}}},
  author = {Rammer, Jorgen},
  year = 2007,
  publisher = {Cambridge University Press}
}

@article{Raston2012,
  title = {Infrared Spectroscopy of {{HOCl}} Embedded in Superfluid Helium Nanodroplets: {{Probing}} the Dynamical Response of the Solvent},
  shorttitle = {Infrared Spectroscopy of {{HOCl}} Embedded in Superfluid Helium Nanodroplets},
  author = {Raston, Paul L. and Kelloway, Donald D. and J{\"a}ger, Wolfgang},
  year = 2012,
  month = jul,
  journal = {The Journal of Chemical Physics},
  volume = {137},
  number = {1},
  pages = {014302},
  issn = {0021-9606},
  doi = {10.1063/1.4731283},
  urldate = {2026-06-08}
}

@article{Redchenko2016,
  title = {Libration of {{Strongly-Oriented Polar Molecules}} inside a {{Superfluid}}},
  author = {Redchenko, E. S. and Lemeshko, Mikhail},
  year = 2016,
  journal = {ChemPhysChem},
  volume = {17},
  number = {22},
  pages = {3649--3654},
  issn = {1439-7641},
  doi = {10.1002/cphc.201601042},
  urldate = {2026-05-28},
  copyright = {\copyright{} 2016 Wiley-VCH Verlag GmbH \& Co. KGaA, Weinheim},
  langid = {english}
}

@article{Rost1992,
  title = {Pendular States and Spectra of Oriented Linear Molecules},
  author = {Rost, J. M. and Griffin, J. C. and Friedrich, B. and Herschbach, D. R.},
  year = 1992,
  month = mar,
  journal = {Physical Review Letters},
  volume = {68},
  number = {9},
  pages = {1299--1302},
  publisher = {American Physical Society},
  doi = {10.1103/PhysRevLett.68.1299},
  urldate = {2026-06-16}
}

@article{Schlesinger2010,
  title = {Dissipative Vibrational Wave Packet Dynamics of Alkali Dimers Attached to Helium Nanodroplets},
  author = {Schlesinger, Martin and Mudrich, Marcel and Stienkemeier, Frank and Strunz, Walter T.},
  year = 2010,
  month = apr,
  journal = {Chemical Physics Letters},
  volume = {490},
  number = {4},
  pages = {245--248},
  issn = {0009-2614},
  doi = {10.1016/j.cplett.2010.03.060},
  urldate = {2026-06-08}
}

@article{Schmidt2015,
  title = {Rotation of {{Quantum Impurities}} in the {{Presence}} of a {{Many-Body Environment}}},
  author = {Schmidt, Richard and Lemeshko, Mikhail},
  year = 2015,
  month = may,
  journal = {Physical Review Letters},
  volume = {114},
  number = {20},
  pages = {203001},
  publisher = {American Physical Society},
  doi = {10.1103/PhysRevLett.114.203001},
  urldate = {2024-02-26}
}

@article{Schmidt2016,
  title = {Deformation of a {{Quantum Many-Particle System}} by a {{Rotating Impurity}}},
  author = {Schmidt, Richard and Lemeshko, Mikhail},
  year = 2016,
  month = feb,
  journal = {Physical Review X},
  volume = {6},
  number = {1},
  pages = {011012},
  publisher = {American Physical Society},
  doi = {10.1103/PhysRevX.6.011012},
  urldate = {2024-02-26}
}

@article{Szalewicz2008,
  title = {Interplay between Theory and Experiment in Investigations of Molecules Embedded in Superfluid Helium Nanodroplets\dag},
  author = {Szalewicz, Krzysztof},
  year = 2008,
  month = apr,
  journal = {International Reviews in Physical Chemistry},
  volume = {27},
  number = {2},
  pages = {273--316},
  publisher = {Taylor \& Francis},
  issn = {0144-235X},
  doi = {10.1080/01442350801933485},
  urldate = {2024-05-16}
}

@article{Thaler2018,
  title = {Femtosecond Photoexcitation Dynamics inside a Quantum Solvent},
  author = {Thaler, Bernhard and Ranftl, Sascha and Heim, Pascal and Cesnik, Stefan and Treiber, Leonhard and Meyer, Ralf and Hauser, Andreas W. and Ernst, Wolfgang E. and Koch, Markus},
  year = 2018,
  month = oct,
  journal = {Nature Communications},
  volume = {9},
  pages = {4006},
  issn = {2041-1723},
  doi = {10.1038/s41467-018-06413-9},
  urldate = {2024-05-16},
  pmcid = {PMC6167364},
  pmid = {30275442}
}

@article{Toennies2004,
  title = {Superfluid {{Helium Droplets}}: {{A Uniquely Cold Nanomatrix}} for {{Molecules}} and {{Molecular Complexes}}},
  shorttitle = {Superfluid {{Helium Droplets}}},
  author = {Toennies, J. Peter and Vilesov, Andrey F.},
  year = 2004,
  journal = {Angewandte Chemie International Edition},
  volume = {43},
  number = {20},
  pages = {2622--2648},
  issn = {1521-3773},
  doi = {10.1002/anie.200300611},
  urldate = {2024-05-16},
  copyright = {Copyright \copyright{} 2004 WILEY-VCH Verlag GmbH \& Co. KGaA, Weinheim},
  langid = {english}
}

@article{Vivanco2025,
  title = {The Strongly Driven {{Fermi}} Polaron},
  author = {Vivanco, Franklin J. and Schuckert, Alexander and Huang, Songtao and Schumacher, Grant L. and Assump{\c c}{\~a}o, Gabriel G. T. and Ji, Yunpeng and Chen, Jianyi and Knap, Michael and Navon, Nir},
  year = 2025,
  month = apr,
  journal = {Nature Physics},
  volume = {21},
  number = {4},
  pages = {564--569},
  issn = {1745-2473, 1745-2481},
  doi = {10.1038/s41567-025-02799-8},
  urldate = {2026-04-23},
  langid = {english}
}

@misc{Wang2025a,
  title = {An Ultraslow Optical Centrifuge with Arbitrarily Low Rotational Acceleration},
  author = {Wang, Kevin and {MacPhail-Bartley}, Ian and Peters, Cameron E. and Milner, Valery},
  year = 2025,
  month = dec,
  number = {arXiv:2512.20568},
  eprint = {2512.20568},
  primaryclass = {physics},
  publisher = {arXiv},
  doi = {10.48550/arXiv.2512.20568},
  urldate = {2026-03-09},
  archiveprefix = {arXiv}
}

@article{Wasak2024,
  title = {Decoherence and {{Momentum Relaxation}} in {{Fermi-Polaron Rabi Dynamics}}: {{A Kinetic Equation Approach}}},
  shorttitle = {Decoherence and {{Momentum Relaxation}} in {{Fermi-Polaron Rabi Dynamics}}},
  author = {Wasak, Tomasz and Sighinolfi, Matteo and Lang, Johannes and Piazza, Francesco and Recati, Alessio},
  year = 2024,
  month = may,
  journal = {Physical Review Letters},
  volume = {132},
  number = {18},
  pages = {183001},
  publisher = {American Physical Society},
  doi = {10.1103/PhysRevLett.132.183001},
  urldate = {2026-06-16}
}

\end{document}


\title{\textsc{Supplemental Material} \\
Crossing the Rotational Sound Barrier in a Quantum Solvent 
}

\author{Baptiste Coquinot}\email{Baptiste.Coquinot@ist.ac.at}
\affiliation{Institute of Science and Technology Austria (ISTA), Am Campus 1, 3400 Klosterneuburg, Austria}

\author{Giacomo Bighin}
\affiliation{University of Zagreb Faculty of Science, Department of Physics, Bijeni\v{c}ka 32, 10000 Zagreb, Croatia}


\author{Mikhail Lemeshko}
\affiliation{Institute of Science and Technology Austria (ISTA), Am Campus 1, 3400 Klosterneuburg, Austria}

\author{Ragheed Alhyder}
\affiliation{Institute of Science and Technology Austria (ISTA), Am Campus 1, 3400 Klosterneuburg, Austria}

\date{\today}

\maketitle


\section{Rotor and drive}

We consider a rotor of rotational constant $B$ coupled to a rotating electric field $\tb{E}$
In the lab frame, the Hamiltonian reads:
\beq \hat\H_{\rm rot}^{\rm LF} = B\hat{\tb{J}}^2  -V_{\rm D}\sin^2(\hat\theta)\cos^2(\hat\phi-\Omega t) \eeq
where $V_{\rm D}=\frac{1}{4}\Delta\alpha |\tb{E}|^2$ denotes the drive amplitude controlled by the molecular polarizability anisotropy $\Delta\alpha$ and the electric-field amplitude $|\tb{E}|$, and $\Omega$ is the rotation frequency. 
In practice, we will use $B=3$ GHz, similar to CS$_2$, and assume the molecule to be symmetric, so that only even rotational numbers contribute.
Going to the co-rotating frame via the operator $\exp(-i\Omega t \hat J_z )$, the Hamiltonian becomes time-independent;
\beq \hat\H_{\rm rot}^{\rm RF} = B\hat{\tb{J}}^2 -\Omega \hat J_z -V_{\rm D}\sin^2(\hat\theta)\cos^2(\hat\phi) \eeq
    
The free rotor eigenstates are $|JM\rangle$ where $J$ is the angular momentum and $M$ its projection onto the $z$-axis. Within this basis:
\beq
    \mathbf{\hat J}^2 |J M\rangle = J(J+1)|J M\rangle, \qquad
    \hat J_z |J M\rangle = M |J M\rangle,
\eeq
However, because the drive breaks rotational symmetry, neither $J$ nor $M$ remain good quantum numbers for the Hamiltonian  $\hat\H_{\rm rot}^{\rm RF} $. Yet, the new eigenstates can be decomposed in this basis:
\beq
   \hat\H_{\rm rot}^{\rm RF} |n\rangle =\E_n^{\rm RF}|n\rangle,  
   \qquad 
|n\rangle    = \sum_{J,M}  C^{(n)}_{JM}
    |JM\rangle,
\eeq
Now, we can estimate the dominant $|JM\rangle$ within the ground state $|0\rangle$:
\beq \left(B\hat{\tb{J}}^2 -\Omega \hat J_z\right)|J M\rangle\approx BJ^2-M\Omega \eeq
Minimizing over $M\leq J$ we expect $M^*\approx J^*$. Then, optimizing over $J$ we expect $J^*\approx \Omega/2B$. Now, the lab frame energy reads
\beq \langle J^*M^*|\hat\H_{\rm rot}^{\rm LF} |J^*M^*\rangle \approx \frac{\Omega^2}{4B}-V_{\rm D} \eeq
Comparing with the naive ground state:
\beq \langle J^*M^*|\hat\H_{\rm rot}^{\rm LF} |J^*M^*\rangle< \langle 00|\hat\H_{\rm rot}^{\rm LF} |00\rangle=0\Longrightarrow \Omega  \lesssim \Omega_{\rm loc}=2\sqrt{B V_{\rm D}} \eeq
which is an upper bound of the rotation frequency at which the rotor stops following the drive.

This frequency also corresponds to frequency of the localisation trap. Indeed, the drive aligns the molecule along the minima of the potential at $\theta^*=\pi/2$ and $\phi^*=0, \pi$. For a strongly trapped molecule,  $\theta= \theta^*+\delta\theta$ and $\phi=\phi^*+\delta\phi$. Expanding the Hamiltonian locally, we find:
\beq \hat\H_{\rm rot}^{\rm RF} = B(p_{\delta\theta}^2+p_{\delta\phi}^2) -\Omega p_{\delta\phi}  + V_{\rm D}(\delta\theta^2 + \delta\phi^2 -1)\eeq
where $p_x$ is the momentum conjugate to the variable $x$. Here, we recognise the Hamiltonian of the harmonic oscillator of harmonic frequency 
\beq \Omega_{\rm loc}=2\sqrt{B V_{\rm D}} \eeq

The rotating-frame rotor Hamiltonian was diagonalized in the free-rotor basis
$|JM\rangle$, truncated at $J\leq J_{\rm max}$. The cutoff was chosen
dynamically according to the classical estimate $J^*\simeq \Omega/(2B)$. For the helium calculations shown
in the main text we used $J_{\rm max}=52$, while for the BEC calculations
the floor value $J_{\rm max}=20$ is sufficient.

We denote $G_0(\omega)$ the bare Green's function of the driven rotor,
\begin{equation}
    \hat G_0^{\rm RF}(\omega)
    =
    \frac{1}{
        \omega - \hat\H_{\rm rot}^{\rm RF}+ i0^+
    }.
\end{equation}
In the eigenbasis of  $\hat\H_{\rm rot}^{\rm RF}$, 
\begin{equation}
  \langle n|  \hat G_0^{\rm RF}(\omega)| m\rangle 
    =
    \frac{\delta_{mn}}{
        \omega -\E_n^{\rm RF}+ i0^+
    }.
\end{equation}

Finally, we describe the eigenstates in the lab frame:
\begin{equation}
    |\psi_n^{\rm LF}(t)\rangle
    =
    e^{-i\Omega t J_z}
    e^{-i\E_n^{\rm RF}t}
    |n\rangle .
\end{equation}
Expanding the rotating-frame eigenstate in the spherical basis,
\begin{equation}
     |\psi_n^{\rm LF}(t)\rangle
    =
    \sum_{J,M}
    C_{JM}^{(n)}
    e^{-i[\E_n^{\rm RF}+\Omega M]t}
    |JM\rangle .
\end{equation}
Thus, each angular momentum component acquires an additional dynamical phase proportional to $M$. The state is stationary in the rotating frame, but corresponds in the laboratory frame to a coherent rotation around the $z$ axis at angular velocity $\Omega$. In terms of spectral function, the peak at $\E_n^{\rm RF}$ in the rotating frame becomes a series of peaks at $\E_n^{\rm RF}+\Omega M$ for all $M$ components in the lab frame.

\section{Bosonic bath}

The rotor is coupled to a bosonic environment composed of modes labelled by linear momentum $k$ and angular momentum $\lambda$ with projection $\mu$ onto the $z$-axis. The total Hamiltonian reads~\cite{Schmidt2015}:
\beq \hat\H^{\rm LF} = \hat\H_{\rm rot}^{\rm LF}+   \sum_{k\lambda\mu}
    \omega_k
    \hat b^\dagger_{k\lambda\mu}
    \hat b_{k\lambda\mu}
  + \sum_{k\lambda\mu}
  U_\lambda(k)
    \left[
        Y_{\lambda\mu}(\hat\theta,\hat\phi)
        \hat b_{k\lambda\mu}
        +
        Y^\dagger_{\lambda\mu}(\hat\theta,\hat\phi)
        \hat b^\dagger_{k\lambda\mu}
    \right]\nonumber
\eeq
where $\w_k$ is the boson excitation energy, $ b^\dagger_{k\lambda\mu}$ is the creation operation and $ Y_{\lambda\mu}$ is the spherical harmonic. Here, $U_\lambda(k)$ is the spherical Bessel-transformed molecule-bath interaction potential:
\begin{equation}
U_\lambda(k)
=
\left[
\frac{8 n k^2 \epsilon_k}
{\omega_k (2\lambda+1)}
\right]^{1/2}
\int_0^\infty dr\,
r^2
U_\lambda(r)
j_\lambda(kr),
\end{equation}
where $n$ is the bath density, $\epsilon_k=k^2/2m$ is the kinetic energy with $m$ the atom mass, and $j_\lambda$ the spherical Bessel function and $U_\lambda(r)$ is the anisotropic component $\lambda$ of the molecule-bath interaction potential. 

In practice, we use only the $\lambda =2$ channel.
We use a van der Waals decay
 \beq 
 U_\lambda(r) =- \left( \frac{C_{6,\lambda}}{r^6}+\frac{C_{8,\lambda}}{r^8}\right)\theta(r-r_\lambda)
 \eeq
 for superfluid helium with parameters extracted from CS$_2$-He interaction~\cite{Farrokhpour2013}:
 \begin{align}
C_{6,2} &= 0.33396 \mathrm{\,GHz.nm^6},\\
C_{8,2} &= 2.6136  \mathrm{\,GHz.nm^8},\\
r_2 &= 0.576\mathrm{\,nm}.
\end{align}
For BECs, we use a Gaussian profile~\cite{Schmidt2015}:
\begin{equation}
U_\lambda(r)
=
\frac{u_\lambda}{(2\pi)^{3/2}}
\exp\!\left(-\frac{r^2}{2r_\lambda^2}\right).
\end{equation}
with
\beq u_2 = 374\tn{ GHz}, \qquad r_2 = 1.04\tn{ nm} \eeq

Going to the co-rotating frame, the Hamiltonian becomes:
\beq \hat\H^{\rm RF} = \hat\H_{\rm rot}^{\rm RF}+   \sum_{k\lambda\mu}
    (\omega_k-\Omega\mu)
    \hat b^\dagger_{k\lambda\mu}
    \hat b_{k\lambda\mu}  + \sum_{k\lambda\mu}
  U_\lambda(k)
    \left[
        Y_{\lambda\mu}(\hat\theta,\hat\phi)
        \hat b_{k\lambda\mu}
        +
        Y^\dagger_{\lambda\mu}(\hat\theta,\hat\phi)
        \hat b^\dagger_{k\lambda\mu}
    \right]\nonumber
\eeq

\section{Renormalised driven rotor}

To second order in the rotor--bath coupling, the impurity self-energy is
\begin{equation}
    \hat\Sigma(\omega)
    =
    \sum_{k\lambda\mu}
    |U_\lambda(k)|^2
    \hat Y_{\lambda\mu}
    \hat  G_0(\omega-\omega_k+\Omega\mu)
     \hat Y^\dagger_{\lambda\mu}
\end{equation}
in the co-rotating frame
and the renormalised Green's function satisfies the Dyson equation
\begin{equation}
     \hat G^{-1}(\omega)
    =
     \hat G_0^{-1}(\omega)
    -
     \hat \Sigma(\omega).
\end{equation}
Thus, the renormalised Green's function in the co-rotating frame reads
\begin{equation}
    \hat G^{\rm RF}(\omega)
    =
    \frac{1}{
        \omega - \hat\H_{\rm rot}^{\rm RF} -\hat \Sigma+ i0^+
    }.
\end{equation}
and the corresponding spectral function is
\begin{equation}
    A(\omega)
    =
    -\frac{1}{\pi}
    \mathrm{Im}\,
    \mathrm{Tr}\,
    \hat G^{\rm RF}(\omega)
\end{equation}

In the weak-coupling regime, the rotor eigenmodes are only weakly renormalised by the bath and the self-energy can therefore be analyzed mode by mode through the diagonal projection
$
\Sigma_n(\omega)
=
\langle n|\hat\Sigma(\omega)|n\rangle.
$
The self-energy mainly contributes on shell, namely for small frequency detuning
$
\Delta\omega
=
\omega-\E_n^{\rm RF}.
$
Thus,
\begin{equation}
    \Sigma_n(\Delta\omega)
    =
    \sum_{k,\mu=-2}^{2}
    \sum_m
    |U_2(k)|^2
    \frac{
        \left|
        \langle m|
        Y_{2\mu}^{\dagger}
        |n\rangle
        \right|^2
    }{
        \Delta\omega
        -\omega_k
        +\Omega\mu
        -(\E_m^{\rm RF}-\E_n^{\rm RF})
        +i0^+
    } 
\end{equation}
with resonance condition:
\beq \Delta\omega
         = \omega_k
        -\Omega\mu
        +(\E_m^{\rm RF}-\E_n^{\rm RF}) \eeq

The imaginary part evaluated on-shell gives the decay rate
\begin{equation}
    \Gamma_n(\Omega)
    =
    -2\,\mathrm{Im}\,
     \Sigma_n(\Omega).
\end{equation}
Thus,
\begin{equation}
\begin{aligned}
    \Gamma_n(\Omega)
    =
    2\pi
    \sum_{k,\mu=-2}^{2}
    \sum_m
    |U_2(k)|^2
    \left|
        \langle m;\Omega|
        Y_{2\mu}^{\dagger}
        |n;\Omega\rangle
    \right|^2
    \delta\!\left(
        \E_n^{\rm RF}(\Omega)
        -\E_m^{\rm RF}(\Omega)
        -\omega_k
        +\Omega\mu
    \right).
\end{aligned}
\end{equation}

The real part of the same expression gives the bath-induced energy shift. At rotational frequency smaller than the resonance condition, all $\mu$-channels contribute to the renormalisation. At larger rotational frequencies, only the $\mu=0$ channel survives and the energy renormalisation becomes weaker.

Numerically, the bath momentum integrals were evaluated by direct quadrature on a uniform
grid of $N_k=2500$ points. The self-energy was evaluated on a frequency
grid of $N_\omega=2500$ points, with a small retarded broadening $\eta$ (or $i0^+$) chosen
to match the frequency resolution. For the helium data in Fig.~2, the frequency
window was $[-100,400]$ GHz, corresponding to $\eta=0.2$ GHz. For the BEC
 data in Fig.~3, the window was $[-25,25]$ GHz, corresponding to $\eta=0.02$ GHz.
 
At each value of $\Omega$, the rotating-frame rotor Hamiltonian was first
diagonalized. The matrix Green's function was constructed in the low-energy
subspace spanned by the lowest $N_{\rm low}=10$ rotor eigenstates (using lab frame energies), while the one-loop self-energy includes virtual transitions through a larger intermediate
manifold of $N_{\rm int}=20$ rotor eigenstates. The spectral function was obtained from the resulting matrix Green's function projected onto the low-energy subspace. 

Quasiparticle energies and linewidths were extracted by fitting the spectral
peaks to a single Lorentzian profile. For each branch we used
\[
A_n(\omega)\simeq A_n^{(0)}
\frac{\gamma_n^2}{(\omega-\E_n)^2+\gamma_n^2},
\]
where $\E_n$ is the fitted quasiparticle energy and the reported linewidth is
$\Gamma_n=2\gamma_n$. The fit was constrained by the frequency resolution
through the lower bound $\gamma_n\geq \eta$, so that linewidths of order
$2\eta$ should be interpreted as resolution-limited. The fitted modes were
ordered by their renormalized energies at each value of $\Omega$.

\bibliography{bibfile}